\documentclass[10pt,twocolumn,prb,showpacs,superscriptaddress,floatfix,float]{revtex4-2}
\usepackage{graphicx}
\usepackage{bm}
\usepackage{amssymb,amsmath,xr}
\usepackage{xcolor,ulem,here,soul}
\def\beq{\begin{equation}}
\def\eeq{\end{equation}}
\def\beqa{\begin{eqnarray}}
\def\eeqa{\end{eqnarray}}
\def\bear{\begin{array}}
\def\eear{\end{array}}

\begin{document}
\title{Josephson diode effect via a non-equilibrium Rashba system}
\author{Michiyasu Mori}
\email{E-mail: mori.michiyasu@jaea.go.jp}
\affiliation{Advanced Science Research Center, Japan Atomic Energy Agency, 
	Tokai, Ibaraki 117-1195, Japan} 

\author{Wataru Koshibae}
\affiliation{RIKEN Center for Emergent Matter Science (CEMS), Wako 351-0198, Japan}
	
\author{Sadamichi Maekawa}
\affiliation{Advanced Science Research Center, Japan Atomic Energy Agency, 
	Tokai, Ibaraki 117-1195, Japan}
\affiliation{RIKEN Center for Emergent Matter Science (CEMS), Wako 351-0198, Japan}

\begin{abstract}  
A non-equilibrium state in a Rashba system under an in-plane magnetic field is identified as the origin of the Josephson diode effect. This state is induced by a current bias—necessary for measuring the current-voltage characteristics—which shifts the Fermi momentum away from equilibrium. This essential mechanism has been overlooked in previous studies. 
This oversight stems from the implicit assumption that the equilibrium-based formulations are sufficient to describe Josephson effect.
We formulate the Josephson coupling energy via the non-equilibrium Rashba system under current bias using a tunneling Hamiltonian, where the Rashba system is modeled as one-dimensional. When the magnetic field is applied perpendicular to the current, the Josephson coupling energy becomes asymmetric, giving rise to the diode effect. The magnitude and sign of this effect depend on the distance between the superconducting electrodes $d$, the in-plane magnetic field, and the spin-orbit coupling strength. 
Our results clarify the microscopic origin of the Josephson diode effect, which can be optimized by tuning $d$.
\end{abstract}

\date{\today}

\maketitle
\section{Introduction}
The DC Josephson effect is measured by applying a DC current to a junction consisting of superconducting electrodes, i.e., Josephson junction~\cite{josephson62,anderson63,tinkham04}.
As the current is increased, a finite voltage suddenly appears across the junction at the critical current $I_c$.
Josephson junctions are classified according to the barrier separating the superconductors.
In junctions with an insulator, Cooper pairs tunnel directly between the superconductors.
In contrast, in junctions separated by a normal metal, the bias current $I_B$ is carried by conduction electrons inside the metallic region.
Although both types of junctions exhibit the Josephson effect and support a dissipationless supercurrent, this distinction becomes crucial when $I_B$ is applied.

While the insulating barrier remains unaffected, the metallic region inevitably becomes a non-equilibrium state due to the applied $I_B$.
The current through the junction is the supercurrent. 
Since the electric current must be continuous through the junction, 
the current flowing in the metallic region is also the supercurrent  
rather than the paramagnetic current driven by an electric field. 
In the metallic region, the phase difference associated with a finite supercurrent can be equivalently represented as an effective vector potential, which reflects the shift of the Fermi momentum in the current-carrying state~\cite{tinkham04,note_gauge}.
The center-of-mass momentum in the Fermi sea becomes finite,
leading to the supercurrent flowing in the metallic region.  
As a result, the electronic state of the metallic region under the current bias is in the non-equilibrium steady state. 
Because $I_c$ corresponds to the critical value of $I_B$, the resulting non-equilibrium state of the normal metal must be taken into account when discussing the Josephson effect via the metallic region.

The Josephson diode effect has attracted significant attention in recent years~\cite{baumgartner22,jeon22,pal22,wu22,kim24,ando24}.
In the conventional Josephson junctions,
the magnitude of $I_c$ is the same regardless of the current direction.
In contrast, the Josephson diode effect refers to a nonreciprocal critical current, whereby the magnitude of $I_c$ depends on the current direction. 
A typical device consists of two superconductors (SCs) coupled via  
a normal metal with a spin-orbit interaction, i.e., Rashba system~\cite{rashba60,rashba84}. 
The in-plane magnetic field is also applied perpendicular to the DC current biasing the junction, i.e., $I_B$. 
When the in-plane magnetic field is fixed and $I_B$ is simultaneously applied to the Rashba system, 
the magnitude of $I_c$ depends on the current direction.
The asymmetry of $I_c$ is reversed by reversing the in-plane magnetic field.

Since the Josephson diode effect emerges under a finite $I_B$ in the Josephson junction separated by a normal metal, i.e., Rashba system, the non-equilibrium electronic state discussed above must be taken into account.
Nevertheless, most previous theories neglected the non-equilibrium nature of the normal metal due to the current bias, and assumed that equilibrium formulations for the Josephson effect remain valid even in the presence of a finite $I_B$~\cite{reynoso08,reynoso12,yokoyama14,zhang22,davydova22,souto22,tanaka22,lu23,hu23,fu24,cayao24,debnath24,soori24,fracasse24,yerin24,ilic24,soori25,debnath25,bhowmik25}. 
However, it is not evident that such assumptions remain valid for the Josephson diode effect.

In this paper, 
we demonstrate that the current-induced non-equilibrium steady state is the key for the Josephson diode effect.
The Josephson coupling energy via the Rashba system under the current bias is formulated using a tunneling Hamiltonian, 
with analytical calculations carried out in a one-dimensional model. 
The Rashba system under the current bias is in the non-equilibrium steady state described by the Fermi momentum shift. 
It is caused by the continuity of the electric current through the junction. 
The Josephson coupling energy is calculated in the fourth order of the tunneling matrix element and reproduces the Josephson diode effect by including $I_B$ in the formulation.
Our analytical results show how the magnitude and the sign of the Josephson diode effect depend on the distance between the superconducting electrodes. 
It will be useful to develop a new guiding principle to design the Josephson diode device.

The rest of this paper is organized as follows.
Section II introduces the model and Hamiltonian.
Section III presents the Josephson coupling energy under a bias current and the corresponding critical current as functions of the spin–orbit interaction, magnetic field, and the distance between the superconducting electrodes.
Section IV discusses the Josephson diode effect, focusing on the asymmetry of the critical current with respect to the magnetic field and electrode distance.
Section V provides a summary and discussion.

\section{Model and Hamiltonian}
\begin{figure*}[htb]
	\centering
	\includegraphics[width=0.95\textwidth]{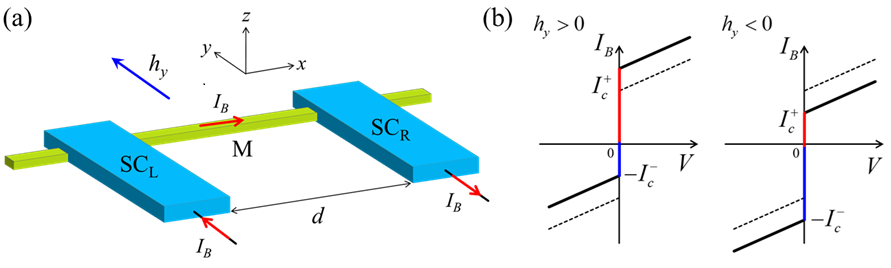}
	\caption{(a) The device geometry of the Josephson junction through the Rashba system (M), which is supposed to be one-dimensional. 
		The two SCs (SC$_{\rm R}$ and SC$_{\rm L}$) are separated by the Rashba system with distance $d$. 
		An external magnetic field $h_y$ is applied in the direction perpendicular to the applied current $I_B$. The signs of $I_B$ and $h_y$ are defined by the Cartesian coordinate system shown in this panel.
        (b) The schematics of the current-voltage curve of the Josephson junction. Due to the Josephson diode effect, the amplitude of critical current $I_c$ in the positive branch ($I_c^+$) colored by red is different from that in the negative one ($I_c^-$) colored by blue. 
        The broken line is the curve without magnetic field, i.e., $h_y=0$. For $h_y>0$ ($h_y<0$), the curve is shifted up (down) from that with $h_y=0$. The shift is reversed by reversing $h_y$. }\label{device}
\end{figure*}

The Josephson junction via the Rashba system (M) is illustrated in Fig.~\ref{device} (a). 
The total Hamiltonian of the junction formed by the SCs ($H_{SC_L}$, $H_{SC_R}$), the Rashba system ($H_M$), and the tunneling between the SCs and the Rashba system ($H_{TL}$, $H_{TR}$) is given by,
\begin{align}
    H &= H_{SC_L} + H_{SC_R} + H_M + H_{TL} + H_{TR}.\label{hamiltonian}
\end{align}
Details of the Hamiltonian are given in the Appendix~\ref{appendix_hamiltonian}.
As shown in Fig.~\ref{device} (a), $I_B$ and the in-plane magnetic field $h_y$ 
are taken to be in the $x$- and $y$-directions, respectively.  
Under $h_y$, the magnitude of $I_c$ depends on the direction of $I_B$, i.e., $I_c^+\ne I_c^-$ (See Fig. 1 (b)).

\begin{figure*}[htb]
	\centering
    \begin{minipage}{0.4\textwidth}
	\includegraphics[width=0.8\textwidth]{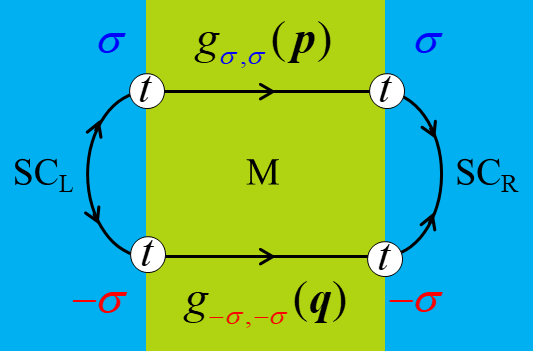}        
    \end{minipage}
    \begin{minipage}{0.4\textwidth}
        \includegraphics[width=0.8\textwidth]{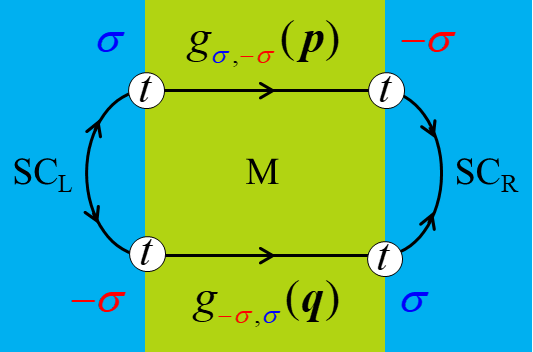}
    \end{minipage}
\caption{Diagrams contributing to the Josephson coupling energy. The solid lines represent the Green's function in each region. Two contributions are shown: non-spin flip (left) and spin flip (right). The Josephson diode effect comes from the process in the right panel. This term appears only when the spin-orbit interaction $\alpha_R$ is there.  }\label{diagram}
\end{figure*}

The Josephson coupling energy $F$ is a function of the phase difference $\phi$ between $SC_L$ and $SC_R$. By taking the derivative of $F$ with respect to $\phi$, the current-phase relation is obtained and the expression of $I_c$ is derived. 
Below, $F$ is calculated in the fourth order of the tunneling matrix element $t$ of $H_{TL}$ and $H_{TR}$ (Appdendix~\ref{appendix_tunnel}). The calculation is diagrammatically shown in Fig.~\ref{diagram}, in which $g_{\sigma,\sigma'}(p)$ is the Green's function of the Rashba system with spins $\sigma$ and $\sigma'$ (Appendix~\ref{appendix_coupling}). 
Figure~\ref{diagram} shows two contributions: non-spin flip (left) and spin flip (right). The latter one is important for the Josephson diode effect. This term appears only when the spin-orbit interaction $\alpha_R$ is finite.  

\begin{figure*}[hbt]
	\centering
	\includegraphics[width=0.95\textwidth]{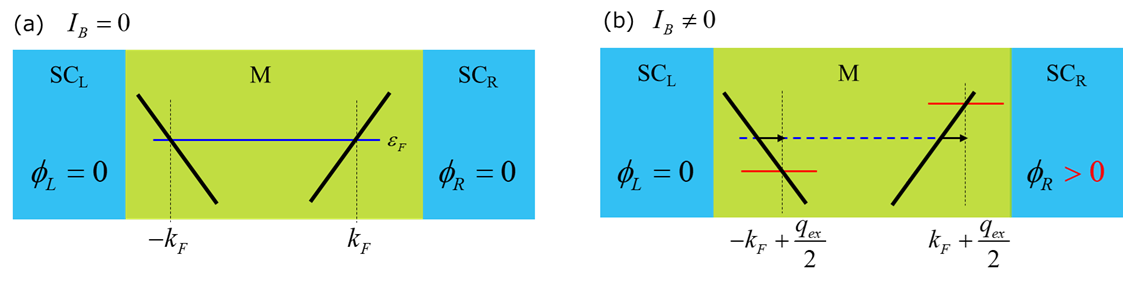}
	\caption{
    The black solid lines indicate the linearized dispersion relation for M in the absence of $\alpha_R$ and $h_y$. 
    In the left panel (a) $I_B=0$, the Fermi energy $\varepsilon_F$ is shown by thin line (blue) and is common to both of left and right branches. In the right panel (b) with $I_B\ne 0$, the Fermi energy in the left branch is different from that in the right branch shown by thin liens (red).}\label{currentbias} 
\end{figure*}

We analytically formulate  $F$
using a one-dimensional model with the linearized dispersion
relation of electrons (see Fig.~\ref{currentbias}). 
Figure~\ref{currentbias} shows how $I_B$ changes the electronic states of M in the absence of $\alpha_R$ and $h_y$.
The current through the junction is the supercurrent relating to $\phi\equiv\phi_L-\phi_R$. 
Since the electric current must be continuous through the junction (red arrows in Fig.~\ref{device} (a)), 
the current flowing in the Rashba system is also the supercurrent induced by $\phi$.  
In the Rashba system, by choosing a different gauge, $\phi$ can be transformed into the vector potential $A\equiv -(\Phi_0/2\pi)\phi/d$ with flux quantum $\Phi_0 (>0)$ and distance between the superconducting electrodes $d$~\cite{note_gauge,tinkham04}.
Such a current-flowing state in M can be expressed by
the Fermi momenta shift as $+k_F \rightarrow +k_F+q_{ex}/2$ and $-k_F \rightarrow -k_F +q_{ex}/2$ as shown in Fig.~\ref{currentbias} (b). 
As a result of the current bias, the electronic state in the Rashba system is changed to a non-equilibrium steady state with the Fermi momentum shift $q_{ex}$. 
\\
\section{Josephson coupling energy and critical current}
By including $I_B$ as a shift of the Fermi momentum $q_{ex}$, $F$ in the fourth order of $t$ is given by, 
\begin{align}
F	&= - t^4\,{\cal U}\,{\cal V} \cos\varphi,\label{fbias}
\end{align}
with gauge invariant phase $\varphi$~\cite{note_gauge,tinkham04} (See Appendix~\ref{appendix_coupling}).
The factor ${\cal U}$ exponentially decays with $d$ for $d/\xi_T \gg 1$ and becomes a power-law decay at low temperatures~\cite{mori07}. 
The other factor ${\cal V}$, on the other hand, 
includes the spin-orbit interaction $\alpha_R$ and is given by,
\begin{widetext}
\begin{align}
{\cal V}
	& =   
	     {\cos\!\left(\frac{\Lambda_- d}{\hbar v_F}\right) 
	      \cos\!\left(\frac{\Lambda_+ d}{\hbar v_F}\right) 
		- \frac{(\gamma  h_z)^2}{\Lambda_- \Lambda_+}
		\sin\!\left(\frac{\Lambda_- d}{\hbar v_F}\right)
		\sin\!\left(\frac{\Lambda_+ d}{\hbar v_F}\right) }
	  +\frac{\lambda_- \lambda_+}{\Lambda_- \Lambda_+}
		\sin\!\left(\frac{\Lambda_- d}{\hbar v_F}\right)
		\sin\!\left(\frac{\Lambda_+ d}{\hbar v_F}\right),\label{biaskernel}
\end{align}
\end{widetext}
\begin{align}
    k_{F\pm} &\equiv k_F \pm q_{ex}/2,\\
	\lambda_\pm &\equiv \alpha_R k_{F\pm} \pm \gamma  h_y,\\
\Lambda_\pm
	& \equiv \sqrt {(\gamma  h_z)^2 + (\lambda_\pm)^2}.
\end{align}
with Fermi momentum $k_F$. 
The parameter $\gamma$ is defined by $\gamma\equiv g\mu_B$ with electron $g$-factor $g$ $(>0)$ and Bohr magneton $\mu_B$. 

Equations~\eqref{fbias} and~\eqref{biaskernel} represent the Josephson coupling energy via the Rashba system under the magnetic field. By taking appropriate limits of Eq.~\eqref{biaskernel}, one can reproduce several known results.
In the following, we set $h_z = 0$, as this corresponds to the setup of the Josephson diode effect shown in Fig.~\ref{device}~(a).
Let us first consider the case without current bias and spin-orbit interaction, i.e., $q_{ex} = \alpha_R =0$.
When $h_y \ne 0$, Eq.~\eqref{biaskernel} reduces to 
\begin{equation}
{\cal V}=\cos\left(\frac{2\gamma h_y}{\hbar v_F}d\right),\label{kernelhy}
\end{equation}
which reproduces the result for the SC/ferromagnet/SC junction~\cite{mori07} realizing the $\pi$-junction~\cite{bulaevskii77,buzdin82}.
As this relation holds for $h_z \ne 0$ and $h_y = 0$, the orientation of the magnetic field does not affect the result.

Next, we turn on the bias current and spin-orbit interaction, i.e., $q_{ex} \ne 0$ and $\alpha_R\ne 0$.
Under this condition, Eq.\eqref{biaskernel} becomes
\begin{align}
{\cal V}
& =
\cos\left[\left(\frac{\alpha_R q_{ex}}{\hbar v_F}+\frac{2\gamma h_y}{\hbar v_F}\right)d\right],\label{biaskernel2}
\end{align}
for $\lambda_{\pm}>0$.
Since Eq.~\eqref{biaskernel2} explicitly depends on $q_{ex}$, i.e., $I_B$, it cannot be obtained within equilibrium theory.
In contrast, Eq.~\eqref{kernelhy} is independent of $q_{ex}$ and can be derived even in equilibrium conditions without the current bias.
It clarifies both why previous studies based on equilibrium states remain valid in certain cases, and 
{\it why incorporating the non-equilibrium state induced by the bias current is essential}.
As we demonstrate below, this expression plays a central role in explaining the Josephson diode effect.
This term emerges when $\alpha_R$, $h_y$, and $I_B$ are simultaneously present.
In this sense, the Josephson diode effect highlights the essential role of the non-equilibrium state induced by the current bias.

The current-phase relation is given by,
\begin{align}
    I_B = \frac{2e}{\hbar}\frac{\partial F}{\partial \varphi}
        = \frac{2e}{\hbar} t^4 {\cal U V} \sin\varphi.\label{cp-relation}  
\end{align}
From Eq.~\eqref{cp-relation}, $I_c$ can be estimated as
\begin{align}
    I_c=(2e/\hbar) t^4{\cal U}{\cal V}\equiv I_{c0} {\cal V},\label{iceq}
\end{align}
where $I_{c0}$ corresponds to $I_c$ of the conventional Josephson junction in the absence of $\alpha_R$ and $h_y$, for which ${\cal V}=1$.
When $I_B$ is increased up to $I_c$, the finite voltage appears.

Due to the current conservation, the same amount of $I_B$ must flow in the Rashba system, in which $I_B$ is generally described by $|I_B|=e$$v_0$$n_e$ with velocity $v_0$ and electron density $n_e$.
Since the electron density contributing to the current is given by $n_e=|q_{ex}|/(2\pi)$, 
$I_B$ is related with $q_{ex}$ by,
\begin{align}
    I_B = -e v_0\frac{q_{ex}}{2\pi}.\label{ib-qex}
\end{align}
Below $v_0=10^4$ m/s is used~\cite{note}. 
As we can see in Eq.~\eqref{biaskernel2}, ${\cal V}$ is a function of $q_{ex}$, i.e., ${\cal V}={\cal V}(q_{ex})$. 
Therefore, Eq.~\eqref{ib-qex} implies that ${\cal V}$ is a function of $I_B$, i.e., ${\cal V}={\cal V}(I_B)$.  
Consequently, $I_c$ must be determined by solving Eq.~\eqref{iceq}, i.e.,  
$I_c=I_{c0}{\cal V}(I_c)$.

In order to estimate the Josephson diode effect, 
$I_c^+$ and $I_c^-$ have to be determined by solving the equations:
\begin{align}
 \frac{I_c^+}{I_{c0}} &= \cos\left[\left(\xi\alpha_R  \frac{I_c^+}{I_{c0}} -\zeta h_y\right)d\right], \label{icplus}\\
 \frac{I_c^-}{I_{c0}} &=\cos\left[\left(\xi\alpha_R  \frac{I_c^-}{I_{c0}} +\zeta h_y\right)d\right], \label{icminus}
\end{align}
with 
$\xi \equiv (1/\hbar v_F)(2\pi/e v_0) I_{c0}$ and 
$\zeta \equiv (2\gamma/\hbar v_F)$ (See Appendix~\ref{appendix_solutions}). 
We take $v_F = 10^4$ m/s and $I_c^0 \sim 1$ nA as reasonable parameters for InAs nanowires~\cite{note,doh05}.
Using these values, we obtain 
$\xi\sim 5.97\times 10^{-5}$ (nm$\cdot$meV\AA)$^{-1}$ 
and 
$\zeta\sim 3.64\times10^{-2}$ (nm$\cdot$T)$^{-1}$. 
The difference between Eqs.~\eqref{icplus} and \eqref{icminus} represents the Josephson diode effect (See Appendix~\ref{appendix_solutions}).

\section{Josephson diode effect: Asymmetry ratio}
\begin{figure}[htb]
	\centering
	\includegraphics[width=0.48\textwidth]{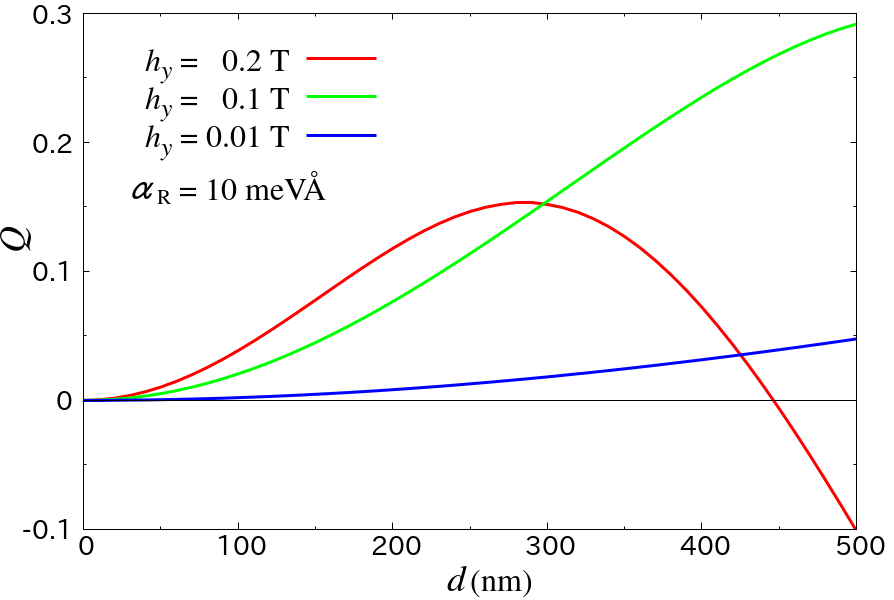}
	\caption{The $d$-dependenced of the assymmetry ratio $Q\equiv (I_c^+-I_c^-)/(I_c^++I_c^-)$ is plotted with $\alpha_R=10$ meV\AA~for 
    $h_y=0.01$ T (blue), 0.1 T (green), and 0.2 T (red).}\label{ddep}
\end{figure}

Numerically solving Eqs.~\eqref{icplus} and \eqref{icminus}, 
the asymmetry ratio $Q$ defined by,
\begin{align}
    Q &\equiv \frac{I_c^+-I_c^-}{I_c^++I_c^-},
\end{align}
is plotted in Fig.~\ref{ddep} as a function of $d$ with $\alpha_R=10$ meV\AA~for 
$h_y=0.01$ T (blue), 0.1 T (green), and 0.2 T (red). 
For small $d$, $Q$ exhibits a quadratic dependence on $d$, i.e., $Q \propto d^2$.
By expanding Eqs.~\eqref{icplus} and \eqref{icminus} up to second order in $d$, we can estimate
\begin{align}
Q &\sim (\xi\alpha_R)\cdot(\zeta h_y)\cdot d^2.\label{q-estimate} 
\end{align}
\def\eeqa{\end{eqnarray}}
Thus, $Q$ increases as $d^2$ and grows linearly with $\alpha_R$ or $h_y$ in the small-$d$ region. 
Although $Q$ increases with $d$, it starts to decrease and becomes negative, e.g., the red line with $h_y=0.2$ T in Fig.~\ref{ddep}. 
The sign change of $Q$ occurs at $I_c^+ = I_c^-$, which corresponds to $\zeta h_y d = \pi n$ with integer $n$ as obtained from Eqs.~\eqref{icplus} and \eqref{icminus}. 
In fact, the red line in Fig.~\ref{ddep} becomes zero at $d=\pi/(\zeta h_y)\sim 446$ nm. 
In this case, the Josephson diode effect can be optimized by tuning of $d$. 
This provides a guiding principle for the Josephson diode device, because $d$ can be controlled in experiments.
It is also important that the Josephson diode effect changes its sign with $d$ for fixed $\alpha_R$ and $h_y$. 
To determine not only the magnitude of the Josephson diode effect, but also its sign, we must take care of $d$ as well.
The $d$-dependence of the Josephson diode effect is the important finding of this study. 

\begin{figure}[htb]
	\centering
	\includegraphics[width=0.48\textwidth]{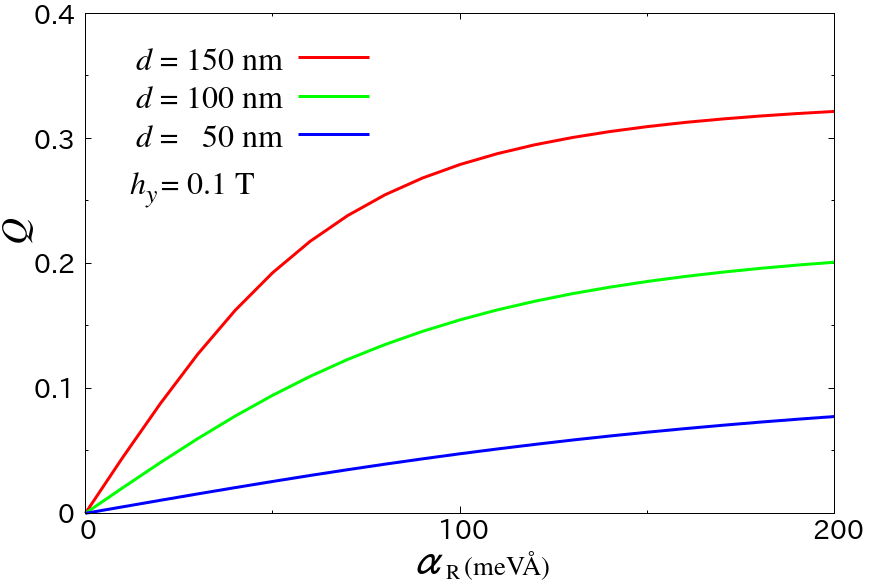}
	\caption{$Q$ is plotted as a function of $\alpha_R$ with $h_y=0.1$ T for $d=50$ nm (blue), 100 nm (green), and 150 nm (red). }\label{adep}
\end{figure}

In Fig.~\ref{adep}, $Q$ is plotted as a function of $\alpha_R$ with $h_y=0.1$ T for $d=50$ nm (blue), 100 nm (green), and 150 nm (red). 
The magnitude of $\alpha_R$ ranges from sub meV\AA~to several hundreds meV\AA~depending on materials and their form, e.g., bulk, film or surface~\cite{lashell96,ast07,ishizaka07}. 
As expected, $Q$ linearly increases with $\alpha_R$ for small $\alpha_R$. 
However, it saturates in the large-$\alpha_R$ region, since higher order terms in $\alpha_R$ are neglected in Eq.~\eqref{q-estimate}. 
Taking account of the next order in $\alpha_R$, $Q$ is approximated as, 
\begin{align}
    Q \sim (\xi\alpha_R)\cdot(\zeta h_y)\cdot d^2\times\left[1-(\xi\alpha_R d)^2\right]. \label{q-estimate-2}
\end{align}
Owing to the last factor in Eq.~\eqref{q-estimate-2}, $Q$ deviates from linear behavior by increasing $\alpha_R$.
Likewise, $Q$ tends to deviate from quadratic behavior as $d$ increases, e.g., the green and red lines in Fig.~\ref{ddep}.

\section{Summary and discussion}
We have formulated the Josephson coupling energy via a non-equilibrium Rashba system under current bias using a tunneling Hamiltonian, where the Rashba system is modeled as one-dimensional. The current bias induces a shift in the Fermi momentum due to current continuity, which plays a central role in the emergence of the Josephson diode effect. Our formulation, incorporating this momentum shift, shows that the diode effect naturally arises in the fourth order of tunneling matrix element. It is worth to note that the higher harmonics of the Josephson coupling energy are not necessary. More importantly, {\it the effect originates from a non-equilibrium response to the bias current}, highlighting a mechanism overlooked in previous studies that were carried out in equilibrium conditions.

When the in-plane magnetic field is applied perpendicular to the bias current, the Josephson coupling energy becomes asymmetric with respect to both the current and the field, leading to the Josephson diode effect. Our results reveal that the diode effect depends sensitively on the distance $d$ between the superconducting electrodes. While tuning the spin-orbit-induced band splitting is generally difficult, $d$ is experimentally controllable, providing a practical route to optimize the diode effect. 

In this study, we have employed a one-dimensional model, which is qualitatively justified for capturing the essential physics of the Josephson diode effect.
The nonreciprocal response originates from the current-induced non-equilibrium steady state, and, in addition, 
the modification of the electronic structure induced by the magnetic field applied perpendicular to the bias current plays an essential role.
These key ingredients are naturally incorporated in our model, allowing us to elucidate the underlying physical mechanism.

Some previous studies have attributed the Josephson diode effect to the anomalous phase shift proposed in Ref.~\cite{buzdin08}, in which the Rashba system is assumed to be the superconducting state. 
We emphasize that the Josephson diode effect discussed in this work originates from the current-induced non-equilibrium steady state in the Rashba system, and does not rely on the anomalous phase shift of the superconducting order parameter.

Finally, we emphasize that although the Josephson diode effect has been widely discussed~\cite{reynoso08,reynoso12,yokoyama14,zhang22,davydova22,souto22,tanaka22,lu23,hu23,fu24,cayao24,debnath24,soori24,fracasse24,yerin24,ilic24,soori25,debnath25}, most previous theories neglected the non-equilibrium nature of the biased junction. By explicitly incorporating the current bias in our formulation, we demonstrate that the Josephson diode effect is an intrinsically non-equilibrium phenomenon driven by the bias current and not requiring higher harmonics. 

\section*{Acknowledgments} 
This work was supported by JSPS Grant Nos.~JP20K03810, JP21H04987, JP23K03291 and the inter-university cooperative research program (No.~202312-CNKXX-0016) of the Center of Neutron Science for Advanced Materials, Institute for Materials Research, Tohoku University. 
WK was supported by CREST Grant No. JPMJCR20T1 from JST.
SM was supported by JSPS Grant No. JP24K00576. 
A part of the computations were performed on supercomputers at the Japan Atomic Energy Agency. 

\begin{appendix}
\section{Hamiltonian}\label{appendix_hamiltonian}
\subsection{Superconductors}\label{appendix_sc}
The Hamiltonian of singlet SC in left (SC$_{\rm L}$, $\lambda$ = L) and right (SC$_{\rm R}$, $\lambda$ = R) electrodes is given by,
\begin{align}
H_{SC_\lambda} 
    &= \sum\limits_{k,\sigma=\pm } 
        \left( H_{\lambda,p} + H_{\lambda,h} + H_{\lambda,i} \right),\\
H_{\lambda,p}	&= 
			v_F\left(k - k_F\right)a_{\lambda k\sigma }^\dag {a_{\lambda k\sigma }} 
		   -v_F\left(k + k_F\right)b_{\lambda k\sigma }^\dag {b_{\lambda k\sigma }},\\
H_{\lambda,h}
	&=v_F\left(k + k_F\right)a_{\lambda\overline{k}\sigma }a_{\lambda\overline{k}\sigma }^\dag 
	       -v_F\left(k - k_F\right)b_{\lambda\overline{k}\sigma}b_{\lambda\overline{k}\sigma }^\dag,\\
H_{\lambda,i}
	&= \sigma\Delta \left(
			a_{\lambda k\sigma }^\dag b_{\lambda\overline{k}\overline{\sigma}}^\dag 
		 +  b_{\lambda k\sigma }^\dag a_{\lambda\overline{k}\overline{\sigma}}^\dag 
			\right) + {\rm H.c.},\label{hamilsc}\\
\Delta&\equiv
	{\cal I} \langle
	a_{\lambda k +} b_{\lambda\overline{k} -} - a_{\lambda k -} b_{\lambda\overline{k} +} 
	\rangle,
\end{align}
with momentum $k$ ($\overline{k}$ $\equiv$ $-k$), Fermi velocity $v_F$, and electrons spin $\sigma$ ($\overline{\sigma}$ $\equiv$ $-\sigma$). 
The singlet superconducting state by interaction ${\cal I}$ is assumed in the both electrodes. 
The electron creation (annihilation) operators around $+k_F$ and $-k_F$ are denoted by 
$a_{\lambda k\sigma}^\dag$ and $b_{\lambda k\sigma}^\dag$ ($a_{\lambda k\sigma}$ and $b_{\lambda k\sigma}$), respectively. 

\subsection{Rashba system}\label{appendix_rashba}
In our theory, the Rashba system is introduced by the following
one-dimensional model,
\begin{align}
{H_M} &= \sum_{k \sigma \sigma'} 
    \Bigl({H_{rp}} + {H_{lp}} + {H_{rh}} + {H_{lh}}\Bigr)\label{rashba}\\
{H_{rp}} &= 
    a_{k\sigma }^\dag 
        \Bigl\{ v_F\left( k - k_F \right)\tau_{0\sigma\sigma'} 
        + \gamma h_z \tau_{3\sigma\sigma'} \nonumber\\
        &\hspace{1cm}  + \left( {\alpha_R {k_F} + {\gamma h_y}} \right)\tau_{2\sigma\sigma'}\Bigr\}
    a_{k\sigma '} \label{H1}\\
{H_{lp}} &= 
    b_{k\sigma }^\dag 
\def\eeqa{\end{eqnarray}}
        \Bigl\{ -v_F\left(k + k_F\right) \tau_{0\sigma\sigma'} 
        + \gamma h_z\tau_{3\sigma \sigma'} \nonumber\\
	&\hspace{1cm} - \left(\alpha_R k_F - \gamma h_y \right)
            \tau _{2\sigma \sigma'}\Bigr\}
    b_{k\sigma'} \label{H2}\\
{H_{rh}} &=
    a_{\overline{k}\sigma}
        \Bigl\{ v_F \left( k + k_F\right) \tau_{0\sigma\sigma'}
			- \gamma h_z \tau _{3\sigma \sigma'} \nonumber\\
	&\hspace{1cm}- \left( \alpha_R k_F + \gamma h_y \right)\tau _{2\sigma \sigma '}        \Bigr\} 
    a_{\overline{k}\sigma'}^\dag \label{H3}\\
{H_{lh}} &=
    b_{\overline{k}\sigma}
        \Bigl\{ -v_F\left(k - k_F \right) \tau_{0\sigma\sigma'}
			- \gamma h_z \tau_{3\sigma \sigma'}\nonumber\\
	&\hspace{1cm} + \left( \alpha_R k_F - \gamma h_y \right)\tau _{2\sigma \sigma'}        y\Bigr\}
    b_{\overline{k}\sigma'}^\dag,\label{H4}
\end{align}
with the Pauli matrix $\tau_n$ ($n=0,1,2,3$), external magnetic field in $y$ ($z$) direction $h_y$ ($h_z$), and a Rashba parameter $\alpha_R$~\cite{rashba60,rashba84}.
The parameter $\gamma$ is defined by $\gamma\equiv g\mu_B$ with electron $g$-factor $g$ and Bohr magneton $\mu_B$. 
The electron creation (annihilation) operators around $+k_F$ and $-k_F$ are denoted by 
$a_{k\sigma}^\dag$ and $b_{k\sigma}^\dag$ ($a_{k\sigma}$ and $b_{k\sigma}$), respectively. \\

\subsection{Tunneling Hamiltonian}\label{appendix_tunnel}
The tunneling Hamiltonian $H_{\rm TL}$ ($H_{\rm TR}$) between SC$_{\rm L}$ (SC$_{\rm R}$) and M is given by,
\begin{align}
H_{T_{\rm L}} &= \hspace{-8pt}\sum\limits_{k,q,\sigma=\pm} 
	\hspace{-5pt}t \left[ 
		\bigl( a_{L,k\sigma }^\dag  + b_{L,k\sigma}^\dag \bigr)
		\bigl( a_{q\sigma }         + b_{q\sigma } \bigr) + H.c. 
	  \right]\nonumber\\
	 & - \hspace{-8pt}\sum\limits_{k,q,\sigma=\pm } 
	\hspace{-5pt}t \left[ 
		\bigl( a_{L, - k\sigma}   + b_{L, - k\sigma} \bigr)
		\bigl( a_{- q\sigma}^\dag + b_{- q\sigma}^\dag \bigr) + H.c. 
	  \right],\\
 {H_{T_{\rm R}}} &=\hspace{-8pt} \sum\limits_{k,q,\sigma=\pm } 
 	\hspace{-5pt}t e^{i(k - q) d}
 	\left[ 
 		 \bigl( a_{R,k\sigma }^\dag + b_{R,k\sigma }^\dag \bigr)
 		 \bigl( a_{q\sigma }        + b_{q\sigma } \bigr) + H.c. 
 	\right]\nonumber\\
      & - \hspace{-8pt}\sum\limits_{k,q,\sigma=\pm } 
 	\hspace{-5pt} t e^{i(k - q)d}
 	\Bigl[ 
 		 \bigl( a_{R, - k\sigma }    + b_{R, - k\sigma } \bigr)
 		 \bigl( a_{- q\sigma }^\dag  + b_{- q\sigma}^\dag \bigr)
 		 + H.c. \bigr] ,
\end{align}
where the distance between SC$_{\rm L}$ and SC$_{\rm R}$ is denoted by $d$. 
The tunneling matrix element $t$ is assumed to be constant.  

\section{Josephson coupling energy}\label{appendix_coupling}
To calculate the Josephson coupling energy, we follow the procedure as in Ref.~\cite{mori07}.
Using Eq.~\eqref{hamiltonian}, the free energy $F$ is given by 
\begin{align}
   F &= -\frac{1}{\beta}\ln{\rm Tr}\left[e^{-\beta H}\right]\nonumber\\
     &= -\frac{1}{\beta}{\rm Tr}\ln \left[-G^{-1} + T \right],\\
   G &=\left[i\omega_n - (H_{SC_{\rm L}} + H_M + H_{SC_{\rm R}})\right]^{-1},\\
   T &= T_{TL} + T_{TR},
\end{align}
where $\omega_n$ is Matsubara frequency of fermion and $\beta=1/k_BT$.
We employ the Matsubara formalism, which is well established for treating electron tunneling (For a standard textbook treatment, see Ref.~\cite{mahan}).
In Ref.~\cite{mori07}, the Green’s function $G$ and the tunneling matrix $T$ can be represented as $6\times6$ matrices.
The matrix dimension originates from two independent degrees of freedom: the particle–hole (Nambu) space and the spatial indices associated with the two superconductors and the ferromagnet.
Consequently, the total dimension is $6 = 2 \times 3$.
In the present model, the corresponding matrices become $24\times24$.
This enlargement arises because the linearized dispersion introduces an additional degree of freedom associated with right- and left-moving electrons, and because the spin degree of freedom must be treated explicitly in the presence of spin–orbit interaction.
Consequently, the total dimension is $24 = 2 ,(\text{particle and hole}) \times 3 ,(\text{SC$_L$, SC$_R$, amd M}) \times 2 ,(\text{left and right movers}) \times 2 ,(\text{spin up and down})$.
In the fourth order of the tunneling matrix element $t$, $F$ is given by,
\begin{align}
    F=-\frac{1}{4}\mathrm{Tr}\left[GTGTGTGT\right],\label{eq:f4} 
\end{align}
among which the terms contributing to the Josephson coupling energy are shown in Fig.~\ref{diagram}. 
As an example, one of terms is given by, 
\begin{equation}
    \sum_{k_1,k_2,k_3,k_4} f_{L}^* (k_1)g_{n\alpha_2,\beta_2}^*(k_2)g_{m\alpha_3,\beta_3}(k_3)f_{R} (k_4), 
\end{equation}
in which  $f_{L}^*(k_1)$ and $f_{R}(k_4)$ are the anomalous Green function in $SC_{\rm L}$ and $SC_{\rm R}$, respectively.
The Greens functions $g_{n\alpha\beta}$  ($n=1 - 4$ and $\alpha,\beta=+,-$) are given by,
\newpage
\begin{widetext}
\begin{align}
	\left(
	\begin{array}{cc}
		g_{1++}({\bm k})& g_{1+-}({\bm k})\\
		g_{1-+}({\bm k})& g_{1--}({\bm k})
	\end{array}
	\right)
	&= \frac{1}{{\left( {vp_+ + {\Lambda _{+} } - i\omega_n } \right)\left( {vp_+ - {\Lambda _{+} } - i\omega_n } \right)}}\left( 
	{\begin{array}{*{20}{c}}
			{i\omega_n  - vp_+ + {\gamma  h_z}}&{ + i\lambda_{+}}\\
			{ - i\lambda_{+}}&{i\omega_n  - vp_+ - {\gamma  h_z}}
	\end{array}} \right),\\
	\left(
	\begin{array}{cc}
		g_{2++}({\bm k})& g_{2+-}({\bm k})\\
		g_{2-+}({\bm k})& g_{2--}({\bm k})
	\end{array}
	\right) 
	&= \frac{1}{{\left( {vq_- + {\Lambda _{-} } + i\omega_n } \right)\left( {vq_- - {\Lambda _{-} } + i\omega_n } \right)}}\left( 
	{\begin{array}{*{20}{c}}
			{i\omega_n  + vq_- + {\gamma  h_z}}&{ - i\lambda_{-}}\\
			{ + i\lambda_{-}}&{i\omega_n  + vq_- - {\gamma  h_z}}
	\end{array}} \right),\\
	\left(
	\begin{array}{cc}
		g_{3++}^*(-{\bm k})& g_{3+-}^*(-{\bm k})\\
		g_{3-+}^*(-{\bm k})& g_{3--}^*(-{\bm k})
	\end{array}
	\right) 
	&= \frac{1}{{\left( {vq_+ + {\Lambda _{+} } - i\omega_n } \right)\left( {vq_+ - {\Lambda _{+} } - i\omega_n } \right)}}\left( 
	{\begin{array}{*{20}{c}}
			{i\omega_n  - vq_+ - {\gamma  h_z}}&{ - i\lambda_{+}}\\
			{ + i\lambda_{+}}&{i\omega_n  - vq_+ + {\gamma  h_z}}
	\end{array}} \right),\\
	\left(
	\begin{array}{cc}
		g_{4++}^*(-{\bm k})& g_{4+-}^*(-{\bm k})\\
		g_{4-+}^*(-{\bm k})& g_{4--}^*(-{\bm k})
	\end{array}
	\right)  
	&= \frac{1}{{\left( {vp_- + {\Lambda _{-} } + i\omega_n } \right)\left( {vp_- - {\Lambda _{-} } + i\omega_n } \right)}}\left( 
	{\begin{array}{*{20}{c}}
			{i\omega_n  + vp_- - {\gamma  h_z}}&{i\lambda_{-}}\\
			{ - i\lambda_{-}}&{i\omega_n  + vp_- + {\gamma  h_z}}
	\end{array}} \right), 
\end{align}
\end{widetext}
where 
\begin{align}
    k_{F\pm} &\equiv k_F \pm q_{ex}/2,\\
    vp_\pm &\equiv {v_F}\left( {k - k_{F\pm}} \right), \\
    vq_\pm &\equiv {v_F}\left( {k + k_{F\pm}} \right), \\
	\lambda_\pm &= \alpha_R k_{F\pm} \pm \gamma  h_y,\\
\Lambda_\pm
	& \equiv \sqrt {(\gamma  h_z)^2 + (\lambda_\pm)^2}.
\end{align}
Evaluating the momentum integrals of SC$_L$ and SC$_R$ and summing over all of them, we obtain the following results,
\begin{widetext}
   \begin{align}
&F	= -\frac{t^4}{4}(\pi\rho_F)^2(k_{\rm B}T)\sum_{n}\frac{ |\Delta|^2 }{\omega_n^2+|\Delta|^2}\Bigl[
	e^{+i\phi}\sum_{k,k'}{\cal A}(k,k')e^{+i\left(k-k'\right)d} + H.c.
	\Bigr],\label{F-4th}\\
&{\cal A}(k,k')
	 \equiv \!\!\sum_{\substack{m=1,2\\m'=3,4}}
	\Bigl\{
	g_{m++}({\bm k'})g_{m'--}^*(-{\bm k}) + g_{m--}({\bm k'})g_{m'++}^*(-{\bm k}) 
    - \bigl[ g_{m-+}({\bm k'})g_{m'-+}^*(-{\bm k}) + g_{m+-}({\bm k'})g_{m'+-}^*(-{\bm k}) \bigr]
	\Bigr\}. 
\end{align} 
\end{widetext}
with 
${\bm k}\equiv(k,i\omega_n)$.
The prime indicates a different momentum and a different frequency. The density of states at Fermi energy are denoted by $\rho_F$. 
Evaluationg the momentum integrals of M, i.e., $k$ and $k'$, one can obtain Eqs.~\eqref{fbias} and~\eqref{biaskernel}, where ${\cal U}$ is given by,
\begin{align}
    {\cal U}=\frac{1}{4}(\pi\rho_F)^2(k_{\rm B}T)\sum_{n}\frac{ |\Delta|^2 e^{-2|\omega_n|d}}{\omega_n^2+|\Delta|^2}.
\end{align}
It is noted that ${\cal U}$ does not contain $\alpha_R$ and decays with $d$~\cite{mori07}.

\section{Derivation and  graphical interpretation of Eqs.~\eqref{icplus} and \eqref{icminus}}\label{appendix_solutions}
The critical current $I_c$ is determined by the amplitude of the sinusoidal term in Eq.~\eqref{cp-relation} and is related to the bias current $I_B$ through the current–phase relation $I_B = I_c \sin \varphi$, with $I_c \equiv I_{c0} \mathcal{V}$.
The factor ${\cal V}$ depends on the bias current $I_B$ and is given by 
\begin{align}
{\cal V}
& =
\cos\left[\left(\frac{\alpha_R q_{ex}}{\hbar v_F}+\frac{2\gamma h_y}{\hbar v_F}\right)d\right],\\
& =
\cos\left[\left(-\frac{\alpha_R}{\hbar v_F}\frac{2\pi}{e v_0}I_B+\frac{2\gamma h_y}{\hbar v_F}\right)d\right],\label{c2}
\end{align}
where  $I_B = -e v_0(q_{ex}/2\pi)$ is used. 
Normalizing $I_B$ by $I_{c0}$, the critical current is expressed as,
\begin{align}
    \frac{I_c}{I_{c0}} &= \cos\left[\left(\xi\alpha_R  \frac{I_B}{I_{c0}} -\zeta h_y\right)d\right],\label{c3}
\end{align}
Taking account of $I_B=I_c\sin\varphi$, the positive and negative $I_c$'s satisfy the relations $I_B=I_c^+$ and $I_B=-I_c^-$, respectively. 
Substituting these relations into Eq.~\eqref{c3}, we obtain Eqs.~\eqref{icplus} and \eqref{icminus}.

\begin{figure}[H]
	\centering
	\includegraphics[width=0.4\textwidth]{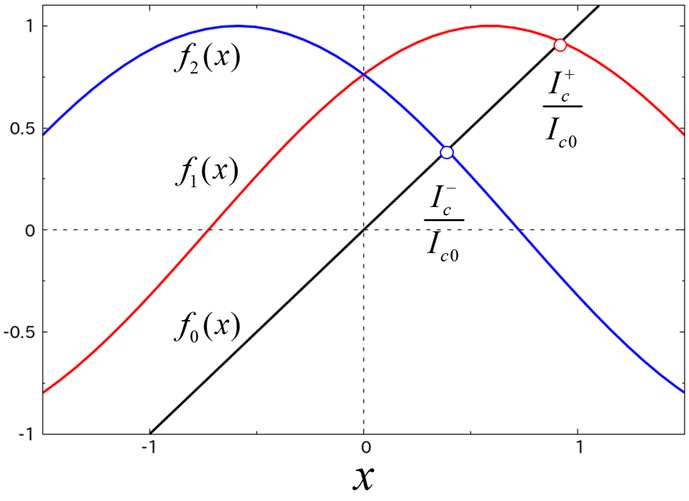}
	\caption{$f_0(x)$, $f_1(x)$, and $f_2(x)$ are plotted for $\alpha_R=100$ meV\AA, $h_y=0.1$ T, $d=200$ nm. The crossing points, red and blue, correspond to the solutions, $I_c^+/I_{c0}$ and $I_c^-/I_{c0}$, respectively. } \label{solutions}
\end{figure}

The solutions of Eqs.~\eqref{icplus} and \eqref{icminus} can be visualized by introducing the following functions:
\begin{align}
    f_0(x) &=x, \\
    f_1(x) &=\cos\left[\left(\xi\alpha_R x - \zeta h_y\right)d\right],\\
    f_2(x) &=\cos\left[\left(\xi\alpha_R x + \zeta h_y\right)d\right], 
\end{align}
where $f_0(x)$ represents $I_c^\pm/I_{c0}$ ($=x$).
The functions $f_1(x)$ and $f_2(x)$ correspond to the right-hand sides of Eqs.~\eqref{icplus} and \eqref{icminus}, respectively.
These functions are plotted for $\alpha_R=100$ meV\AA, $h_y=0.1$ T, $d=200$ nm in Fig.~\ref{solutions}.
\noindent
The crossing points of $f_0(x)$ with $f_1(x)$ and $f_2(x)$ correspond to the solutions for $I_c^+/I_{c0}$ and $I_c^-/I_{c0}$, respectively.
If any one of the three parameters—$d$, $\alpha_R$, or $h_y$—is zero, the functions $f_1(x)$ and $f_2(x)$ become degenerate, and the corresponding solutions also coincide, i.e., $I_c^+ = I_c^- = I_{c0}$. Therefore, the three factors, $\alpha_R$, $h_y$ and $d$, are the necessary condition to observe the Josephson diode effect.  
It is noted that the Josephson diode effect depends on $d$ as well as $h_y$ and $\alpha_R$. 
This means that {\it the propagation of Cooper pairs in the Rashba system is essential}.

The relative magnitude of $I_c^+$ and $I_c^-$ is determined by the relative sign between $\alpha_R$ and $h_y$, namely whether $\alpha_R\cdot h_y$ is positive or negative. 
Since the $I_c^+ - I_c^-$ reflects the sign of the product $\alpha_R \cdot h_y$, one can use this relation to extract the sign of $\alpha_R$.
In other words, once the directions of $I_B$ and $h_y$ are fixed, the sign of $\alpha_R$ can be identified.
This is also the important conclusion drawn from Eqs.~\eqref{icplus} and \eqref{icminus}. 

 \end{appendix}


\begin{thebibliography}{00}
\bibitem{note_gauge}
The gauge-invariant phase $\varphi$ is defined as 
$\varphi \equiv \phi -(2\pi/\Phi_0)\int ds A$, where $\Phi_0$ is the flux quantum~\cite{tinkham04}.
The integral runs from one superconductor to the other across the Rashba system.
When $A=0$ is chosen, the current–phase relation reduces to $I = I_c \sin\phi$.
Alternatively, 
by choosing a gauge in which the superconducting order parameter is real,
a finite vector potential 
satisfying $\varphi = -(2\pi/\Phi_0)\int ds A$
is introduced in the Rashba system.
Note that the term $-(2\pi/\Phi_0)\int ds A$ corresponds to $-q_{ex}d$ under a current bias.
These two descriptions are equivalent, representing different gauge choices, and they describe the same physical situation from different viewpoints\cite{tinkham04}.
\bibitem{note}
In an InAs quantum wire~\cite{doh05}, $I_c$ is on the order of nA and an electron density $n_e$ is estimated to be $10^{18}$ cm$^{-3}$.
Assuming the parabolic dispersion with the bare electron mass, we estimate the Fermi velocity as $v_F=10^4$ m/s.
We set $v_0 = v_F$ in this paper.
\end{thebibliography}
\end{document}